\documentclass[a4paper,11pt]{article}
\usepackage[margin=2cm]{geometry}
\usepackage{tikz,amsmath, amssymb,bm,color, amsthm}
\usetikzlibrary{positioning, calc,chains,fit,shapes}
\usepackage{enumerate}
\usepackage{comment}
\usepackage{tikz}
\usepackage{graphics}
\usepackage{longtable}
\usepackage{mdframed}
\usepackage{caption}
\usepackage{subcaption}
\usepackage{slashbox}
\usepackage{url}
\usepackage{framed}
\usepackage{array}
\usepackage{tabu}
\usepackage{lscape}
\usepackage{multirow}
\usepackage{ulem}
\usepackage{multicol}
\usepackage{placeins}

\usepackage[noadjust]{cite}
\usepackage[inline]{enumitem}
\usepackage[utf8]{inputenc}
\usepackage[T1]{fontenc}

\usepackage{authblk}

\newlist{mylist}{enumerate*}{1}
\setlist[mylist]{label=(\roman*)}

\mdfsetup{skipabove=2pt,skipbelow=2pt}

\newtheorem{theorem}{Theorem}[section]
\newtheorem{lemma}[theorem]{Lemma}
\newtheorem{observation}[theorem]{Observation}

\newtheorem{proposition}[theorem]{Proposition}

\newcommand{\pname}[1]{\textnormal{\textsc{#1}}}

\newcounter{rowcntr}[table]
\renewcommand{\therowcntr}{\thetable.\arabic{rowcntr}}

\newcolumntype{N}{>{\refstepcounter{rowcntr}\therowcntr}c}

\AtBeginEnvironment{longtabu}{\setcounter{rowcntr}{0}}

\newcounter{rowcntra}[table]
\renewcommand{\therowcntra}{\arabic{rowcntra}}

\newcolumntype{M}{>{\refstepcounter{rowcntra}\therowcntra}c}

\AtBeginEnvironment{tabular}{\setcounter{rowcntra}{0}}

\newcommand{\SCG}{\pname{SC-to-$\mathcal{G}$}}

\newcommand{\SCT}{\pname{SC-to-}}

\newcommand{\YES}{YES}
\newcommand{\NO}{NO}

\usetikzlibrary{fit}
\usetikzlibrary{
  graphs,
  graphs.standard
}

\usetikzlibrary{decorations.shapes}

\author[1]{Dhanyamol Antony}
\author[2]{Sagartanu Pal}
\author[2]{R. B. Sandeep}

\affil[1]{Indian Institute of Science, Bangalore, India

\texttt{dhanyamola@iisc.ac.in}}

\affil[2]{Indian Institute of Technology Dharwad, India

\texttt{\{183061001,sandeeprb\}@iitdh.ac.in}}

\title{Algorithms for subgraph complementation to some classes of graphs\footnote{Partially supported by SERB Core Research Grant CRG/2022/006770: ``Bridging Quantum Physics with Theoretical Computer Science and Graph Theory'' under Prof. Sunil Chandran Leela while Dhanyamol Antony was a Postdoc at IISC Bangalore, and by SERB MATRICS Grant MTR/2022/000692: ``Algorithmic study on hereditary graph properties''.}}
\date{}
\begin{document}
\maketitle
\begin{abstract}
    For a class $\mathcal{G}$ of graphs, the objective
    of \textsc{Subgraph Complementation to} $\mathcal{G}$ is to find whether there exists a subset $S$ of vertices of the input graph $G$ such that modifying $G$ by complementing the subgraph induced by $S$ results in a graph in $\mathcal{G}$. We 
    obtain a polynomial-time algorithm for the problem
    when $\mathcal{G}$ is the class of graphs with minimum degree at least $k$, for a constant $k$,
    answering an open problem by Fomin et al. (Algorithmica, 2020). When $\mathcal{G}$
    is the class of graphs without any induced copies of the star graph on $t+1$ vertices (for any constant $t\geq 3$) and diamond, we obtain a polynomial-time algorithm for the problem. 
    This is in contrast with a result by Antony et al. (Algorithmica, 2022) that the problem is NP-complete
    and cannot be solved in subexponential-time (assuming the Exponential Time Hypothesis) when $\mathcal{G}$ is the class of graphs without any induced copies of the star graph on $t+1$ vertices, for every constant $t\geq 5$.
\end{abstract}

\section{Introduction}

Complementation is a very fundamental graph operation and 
modifying a graph by complementing an induced subgraph to satisfy certain properties is a natural algorithmic problem on graphs. The operation of complementing an induced subgraph, known as subgraph complementation, is introduced by Kami{\'n}ski et al.~\cite{DBLP:journals/dam/KaminskiLM09} in connection with clique-width of graphs. For a class $\mathcal{G}$ of graphs, the objective of \textsc{Subgraph Complementation to} $\mathcal{G}$ is 
to find whether there exists a subset $S$ of the vertices of the input graph $G$ such that complementing the subgraph induced by $S$ in $G$ results in a graph in $\mathcal{G}$.  Fomin et al.~\cite{DBLP:journals/algorithmica/FominGST20} studied this problem on various classes $\mathcal{G}$ of graphs.
They obtained that the problem can be solved in 
polynomial-time when $\mathcal{G}$ is bipartite, d-degenerate, or co-graphs. In addition to this, they 
proved that the problem is NP-complete when $\mathcal{G}$ is the class of all regular graphs.
Antony et al.~\cite{DBLP:journals/algorithmica/AntonyGPSSS22} studied this problem when $\mathcal{G}$ is the class of $H$-free graphs (graphs without any induced copies of $H$). They proved that the problem
is polynomial-time solvable when $H$ is a complete graph on $t$ vertices. They also proved that the problem is NP-complete when $H$ is a star graph on at least 6 vertices or a path or a cycle on at least 7 vertices. Later Antony et al.~\cite{DBLP:conf/latin/AntonyPSS22} proved that
the problem is polynomial-time solvable when $H$ is paw, and NP-complete when $H$ is a tree, except for 41 trees of at most 13 vertices. It has been proved~\cite{DBLP:journals/algorithmica/AntonyGPSSS22,DBLP:conf/latin/AntonyPSS22}
that none of these hard problems admit subexponential-time algorithms (algorithms running in time $2^{o(n)}$), assuming the Exponential Time Hypothesis.

Fomin et al.~\cite{DBLP:journals/algorithmica/FominGST20} 
proved that the problem is polynomial-time solvable not only when $\mathcal{G}$ is the class of $d$-degenerate graphs but also when $\mathcal{G}$ is any subclass of $d$-degenerate graphs recognizable in polynomial-time. 
This implies that the problem is polynomial-time solvable when $\mathcal{G}$ is the class of $r$-regular
graphs or the class of graphs with maximum degree at most $r$ (for any constant $r$). They asked whether the problem can be solved in polynomial-time when $\mathcal{G}$ is the class of graphs with minimum
degree at least $r$, for a constant $r$. We resolve this positively and obtain a stronger result - a simple quadratic kernel for the following parameterized problem: Given a graph $G$ and an
integer $k$, find whether $G$ can be transformed into
a graph with minimum degree at least $k$ by subgraph
complementation (here the parameter is $k$).  
The result follows from an observation that if $G$ has 
more than $2k^2-2$ vertices, then it is a yes-instance of the problem.

When $\mathcal{G}$ is the class of graphs without any 
induced copies of the star graph on $t+1$ vertices (for any fixed $t\geq 3$) and the diamond (\begin{tikzpicture}[myv/.style={circle, draw, inner sep=1.5pt}]
    \node (o) at (0,0) {};
    \node[myv] (v1) at (0,0.15) {};
    \node[myv] (v2) at (0,-0.15) {};
    \node[myv] (v3) at (-0.3,0) {};
    \node[myv] (v4) at (0.3,0) {};

    \draw (v1) -- (v2);
    \draw (v1) -- (v3);
    \draw (v1) -- (v4);
    \draw (v2) -- (v3);
    \draw (v2) -- (v4);

\end{tikzpicture}), we obtain a polynomial-time algorithm. 
When $t=3$ this graph class is known as linear domino and is the class of line graphs of triangle-free graphs. Cygan et al.~\cite{DBLP:journals/mst/CyganPPLW17} have studied the polynomial kernelization of edge deletion problem for this target graph class. When $t=4$, the graph class is the line graphs of linear hypergraphs of rank 3. 
The technique that we use is similar to that given in \cite{DBLP:journals/algorithmica/AntonyGPSSS22} and \cite{DBLP:conf/latin/AntonyPSS22} for obtaining 
polynomial-time algorithms when $\mathcal{G}$ is $H$-free, for $H$ being a complete graph on $t$ vertices or a paw. Our result is in contrast with the result
by Antony et al.~\cite{DBLP:journals/algorithmica/AntonyGPSSS22} that the problem is NP-complete and cannot be solved in subexponential-time (assuming the Exponential Time Hypothesis) when $H$ is a star graph on $t+1$ vertices, for every constant $t\geq 5$.

\subsubsection*{Preliminaries}
A diamond is the graph \begin{tikzpicture}[myv/.style={circle, draw, inner sep=1.5pt}]
    \node (o) at (0,0) {};
    \node[myv] (v1) at (0,0.15) {};
    \node[myv] (v2) at (0,-0.15) {};
    \node[myv] (v3) at (-0.3,0) {};
    \node[myv] (v4) at (0.3,0) {};

    \draw (v1) -- (v2);
    \draw (v1) -- (v3);
    \draw (v1) -- (v4);
    \draw (v2) -- (v3);
    \draw (v2) -- (v4);

\end{tikzpicture}, and 
a star graph on $t+1$ vertices, denoted by $K_{1,t}$,
is the tree with $t$ degree-1 vertices and one degree-$t$ vertex. The degree-$t$ vertex of a star is known as the center of the star.
 For example, $K_{1,3}$, also known as a claw,
is the graph \begin{tikzpicture}[myv/.style={circle, draw, inner sep=1.5pt}]
    \node[myv] (o) at (0,0) {};
    \node[myv] (v1) at (-0.3,-0.3) {};
    \node[myv] (v2) at (0,-0.3) {};
    \node[myv] (v3) at (0.3,-0.3) {};
    \draw (o) -- (v1);
    \draw (o) -- (v2);
    \draw (o) -- (v3);
\end{tikzpicture}
. A complete graph
on $t$ vertices is denoted by $K_t$. By $\overline{G}$ we denote the complement graph of $G$. The open neighborhood and closed neighborhood of a vertex $v$ are denoted by $N(v)$
and $N[v]$ respectively. The underlying graph will be evident from the context. 
For a subset $S$ of vertices of $G$, by $G[S]$ we 
denote the graph induced by $S$ in $G$.
For a given graph $G$ and a set  $S\subseteq V(G)$, we define the graph
$G\oplus S$ as the graph obtained from $G$ by complementing the subgraph induced by $S$, i.e., 
an edge $uv$ is in $G\oplus S$ if and only if 
$uv$ is a nonedge in $G$ and $u,v\in S$, or $uv$ is 
an edge in $G$ and $\{u,v\} \setminus S \neq \emptyset$.
The operation is called subgraph complementation.
Let $\mathcal{H}$ be a set of graphs. We say that a graph $G$ is $\mathcal{H}$-free if $G$ does not have any induced copies of any of the graphs in $\mathcal{H}$. If $\mathcal{H}=\{H\}$, then we say that $G$ is $H$-free. The general definition of the problem that we deal with is given below. 
\begin{mdframed}
  \textbf{\SCT{$\mathcal{G}$}\ }:  
  Given a graph $G$,  find whether there is a set $S\subseteq V(G)$ such that $G\oplus S \in \mathcal{G}$ .
\end{mdframed}

In a parameterized problem, apart from the usual input, there is an additional integer input known as the parameter. A graph problem is fixed-parameter tractable (FPT)
if it can be solved in time $f(k)n^{O(1)}$, where $n$
is the number of vertices and $f(k)$ is any computable function. A parameterized problem admits a kernel if there is a polynomial-time algorithm which takes as input an instance $(I',k')$ of the problem and outputs
an instance $(I,k)$ of the same problem so that
$|I|,k\leq f(k)$ for some computable function $f(k)$, and
$(I',k')$ is a yes-instance if and only if $(I,k)$
is a yes-instance (here, $k'$ and $k$ are the parameters). A kernel is a polynomial kernel
if $f(k)$ is a polynomial function. It is known that
a problem admits an FPT algorithm if and only if it 
admits a kernel. An FPT algorithm implies that there 
is a polynomial-time algorithm to solve the problem
when the parameter is a constant. We refer to the book~\cite{DBLP:books/sp/CyganFKLMPPS15} for further exposition on these topics.

\section{Algorithms}
We obtain our results in this section.
Let $\mathcal{G}_k$ be the class of graphs with minimum degree at least $k$. We prove that a no-instance of \SCT{$\mathcal{G}_k$} cannot be very large.
\begin{lemma}
    \label{lem:min}
    Let $G$ be a graph with more than $2k^2-2$ vertices.
    Then $G$ is a yes-instance of \SCT{$\mathcal{G}_k$}.
\end{lemma}
\begin{proof}
    Let $M$ be the set of vertices in $G$ with degree less than $k$. Clearly, $M\subseteq S$ for every solution $S$ (i.e., $G\oplus S\in \mathcal{G}_k$).
    Let $|M|=m$. 
 Let $M'$ be the set of vertices in $V(G)\setminus M$
    adjacent to at least one vertex in $M$. As each vertex in $M$ has degree at most $k-1$, we obtain that $|M'|\leq m(k-1)$.  
    
    Let $M''=V(G)\setminus (M\cup M')$. Let $X$ be the set of vertices in $M''$ having degree at least 
    $2k-m-1$ in $G$. If $|X|\geq k$, then $G\oplus (M\cup X')\in \mathcal{G}_k$, where $X'$ is any subset of $k$ vertices of $X$ - note that degree of every vertex in $X'$ is at least $(2k-m-1)+m-(k-1)=k$, in $G\oplus (M\cup X')$.
    Therefore, assume that
    $|X|\leq k-1$. Every vertex in $M''\setminus X$
    has degree at most $2k-m-2$ in $G$. Then, every
    maximal independent set in $M''\setminus X$ has 
    size at least $|M''\setminus X|/(2k-m-1)$. Therefore,
    if $|M''\setminus X|\geq k(2k-m-1)$, then for any maximal independent set $I$ of $M''\setminus X$, $G\oplus (M\cup I)\in \mathcal{G}_k$. Hence assume that $|M''\setminus X|\leq k(2k-m-1)-1$. Therefore, if $G$ is a no-instance of \SCT{$\mathcal{G}_k$}, then the number of vertices in $G$ is  
    at most $|M|+|M'|+|X|+|M''\setminus X|\leq m+m(k-1)+(k-1)+k(2k-m-1)-1=2k^2-2$. 
\end{proof}

Lemma~\ref{lem:min} gives a polynomial-time algorithm for the problem: If $G$ has more than $2k^2-2$ vertices, then return \YES, and do an exhaustive search for a solution otherwise. 
Lemma~\ref{lem:min} also gives a simple quadratic kernel for the problem parameterized by $k$: For an input $(G,k)$
if $G$ has more than $2k^2-2$ vertices, then return
a trivial yes-instance, and return the same instance otherwise. By a result from~\cite{DBLP:journals/algorithmica/AntonyGPSSS22}, \SCT{$\mathcal{G}$} and \SCT{$\overline{\mathcal{G}}$}
are polynomially equivalent. Therefore, we obtain 
a polynomial-time algorithm for \SCG\ when $\mathcal{G}$ is the class of graphs with maximum degree at most $n-k$, for a constant $k$. It also 
implies a quadratic kernel for the problem parameterized by $k$. It remains open whether the following problem is NP-complete: Given a graph $G$ and an integer $k$, find whether $G$ can be subgraph complemented to a graph with minimum degree at least $k$. We note that, the problem is NP-complete if the objective is to make the input graph $k$-regular~\cite{DBLP:journals/algorithmica/FominGST20}.

\subsubsection*{Destroying stars and diamonds}
Let $\mathcal{G}$ be the class of $\{K_{1,t}, \text{diamond}\}$-free graphs, for any fixed $t\geq 3$. We give a polynomial-time algorithm for \SCG.
The concept of $(p, q)$-split graphs was introduced by Gy\'{a}rf\'{a}s~\cite{DBLP:journals/jct/Gyarfas98}.
For $p\geq 1$, and  $q\geq 1$, if the vertices of a graph $G$ can be partitioned into two sets
$P$ and $Q$ in such a way that
the clique number of $G[P]$ and the independence number of $G[Q]$
are at most $p$ and $q$ respectively (i.e., $G[P]$ is $K_{p+1}$-free and $G[Q]$ is $(q+1)K_1$-free), 
then $G$ is called a $(p,q)$-split graph and $(P,Q)$ is a $(p,q)$-split partition of $G$.

\begin{proposition}[\cite{DBLP:conf/fsttcs/KolayP15, kolay2015parameterized, DBLP:journals/algorithmica/AntonyGPSSS22}]
\label{pro:split-part}
 For any fixed constants $p\geq 1$ and $q\geq 1$, 
 recognizing a $(p,q)$-split graph and obtaining all $(p,q)$-split partitions of a $(p,q)$-split graph 
 can be done in polynomial-time.
\end{proposition}

\begin{mdframed}
\textbf{Algorithm for \SCG}, where $\mathcal{G}$ is $\{K_{1,t},diamond\}$-free graphs, for any constant $t\geq 3$.\\
Input: A graph $G$.\\
Output: If $G$ is a yes-instance of \SCG, 
then returns \YES; otherwise returns \NO.
\begin{description}

\item [Step 1]: Let $S$ be the set of all degree-2 vertices of all the induced diamonds in $G$. If $G\oplus S\in \mathcal{G}$, then return \YES. 
\item [Step 2]: Let $r$ be the center of any induced 
$K_{1,t}$ in $G$ and let $I$ be the set of isolated
vertices in the subgraph induced by $N(r)$ in $G$.
For every subset $S\subseteq I$ such that $|S|\geq |I|-t+2$, if $G\oplus S\in \mathcal{G}$, then return \YES.
\item[Step 3]: For every edge $uv$ in $G$, do the following:
\begin{enumerate}
    \item If $N(u)\setminus N[v]$ or $N(v)\setminus N[u]$ does not induce a 
    $(t-1,t-1)$-split graph, then continue with Step 3.
    \item Compute $L(u\overline{v})$, the list of all $(t-1,t-1)$-split partitions of the graph induced by $N(u)\setminus N[v]$.
    \item Compute $L(\overline{u}v)$, the list of all $(t-1,t-1)$-split partitions of the graph induced by $N(v)\setminus N[u]$.
    \item Compute $L(uv)$, the list of all partitions of 
    the graph induced by $N(u)\cap N(v)$ into an independent set of size at most $t-1$ and the rest.
    \item For every $(S_1,T_1)\in L(u\overline{v})$, 
    for every $(S_2,T_2)\in L(\overline{u}v)$, for 
    every $(S_3,T_3)\in L(uv)$, do the following:
    \begin{enumerate}
        \item Let $S=S_1\cup S_2\cup S_3\cup \{u,v\}$.
        If $G\oplus S\in \mathcal{G}$, return \YES.
        \item For every vertex $w\in \overline{N[u]}\cap \overline{N[v]}$,
        let $S= S_1\cup S_2\cup S_3\cup \{u,v,w\}$. If $G\oplus S\in \mathcal{G}$, return \YES.
        \item For every edge $xy$ in the graph induced 
        by $\overline{N[u]}\cap \overline{N[v]}$, if the graph induced by
        $J=N[x]\cap N[y]\cap \overline{N[u]}\cap \overline{N[v]}$ is not a split graph then 
        continue with the current step. Otherwise,
        for every split partition $(S_4,T_4)$ of the graph
        induced by $J$, let $S= S_1\cup S_2\cup S_3\cup S_4\cup \{u,v\}$. If $G\oplus S\in \mathcal{G}$, then return \YES. 
    \end{enumerate}
\end{enumerate}
\item[Step 4]: Return \NO.
\end{description}
\end{mdframed}
\vspace{0.1cm}
Lemma~\ref{lem:is-diamond} and \ref{lem:is-k1t} deals with the case when $G$ is a yes-instance having a solution which is an independent set, the case handled in Step 1 and 2 of the algorithm.
\begin{lemma}
    \label{lem:is-diamond}
    Assume that $G$ is not diamond-free.
    Let $S\subseteq V(G)$ such that $G\oplus S\in \mathcal{G}$ and $S$ is an independent set. Then $S$
    is the set of all degree-2 vertices of all the induced diamonds in $G$.
\end{lemma}
\begin{proof}
    Since $S$ is an independent set and $G\oplus S\in \mathcal{G}$, both the degree-2 vertices of every induced diamond in $G$ must be in $S$. Assume 
    for a contradiction that $S$ has a vertex $v$
    which is not a degree-2 vertex of any of the induced diamonds in $G$. Let $D=\{d_1,d_2,d_3,d_4\}$
    induces a diamond in $G$, where $d_1$ and $d_2$
    are the degree-2 vertices of the diamond. 
    Clearly, $S\cap D=\{d_1,d_2\}$. We know that $v\neq d_1$ and $v\neq d_2$. If $v$ is not adjacent to $d_3$ in $G$, then $\{v,d_1,d_2,d_3\}$ induces 
    a diamond in $G\oplus S$, which is a contradiction.
    Therefore, $v$ is adjacent to $d_3$. 
    Similarly, $v$ is adjacent to $d_4$. 
   Then $\{v,d_1,d_3,d_4\}$ induced a diamond in $G$,
    where $v$ and $d_1$ are the degree-2 vertices, which is a contradiction.
\end{proof}

\begin{lemma}
    \label{lem:is-k1t}
    Assume that $G$ has no induced diamond but has 
    at least one induced $K_{1,t}$. Let $S\subseteq V(G)$ such that $G\oplus S\in \mathcal{G}$
    and $S$ is an independent set. Let $r$
    be the center of any induced $K_{1,t}$ in $G$.
    Let $I$ be the set of isolated vertices in the
    subgraph induced by $N(r)$ in $G$. Then $S\subseteq I$ and $|S|\geq |I|-t+2$.
\end{lemma}
\begin{proof}
    If $r\in S$, then none of the vertices in $N(r)$
    is in $S$ - recall that $S$ is an independent set.
    But then, none of the induced $K_{1,t}$ centered at
    $r$ is destroyed in $G\oplus S$.
    Therefore, $r\notin S$.
    Since $G$ is diamond-free, $N(r)$
    induces a cluster (graph with no induced path of length 3) $J$ in $G$.
    Since $r$ is the center 
    of an induced $K_{1,t}$ in $G$, there are at 
    least $t$ cliques in $J$.
    Since $G\oplus S$ is $K_{1,t}$-free,
    $S$ must contain all vertices of at least two
    cliques in $J$. Since $S$ is an independent set,
    $S$ contains at least two isolated vertices, say $s_1$ and $s_2$, in $J$. First we prove that
    $S\subseteq N(r)$. 
    For a contradiction, assume that there is a vertex
    $v\in S$ such that $v$ is not adjacent to $r$. 
    Then $\{v,s_1,s_2,r\}$ induces a diamond
    in $G\oplus S$, which is a contradiction.
    Therefore, $S\subseteq N(r)$. Next we prove that
    $S\subseteq I$. For a contradiction, assume that
    there is a vertex $v\in S\setminus I$. Then $v$
    is part of a clique $J'$ of size at least 2 in $J$.
    Let $v'$ be any other vertex in $J'$. Since 
    $S$ is an independent set, $v'\notin S$. 
    Then $\{v,v',s_1,r\}$ induces a diamond in $G\oplus S$, which is a contradiction. Therefore, $S\subseteq I$. If $|S|<|I|-t+2$, then there is 
    a $K_{1,t}$ centered at $r$ in $G\oplus S$, which 
    is a contradiction.
\end{proof}

Let $G$ be a yes-instance of \SCG.
Let $S\subseteq V(G)$ be such that $|S|\geq 2$, $G\oplus S\in \mathcal{G}$, and $S$ be not an independent set.
Let $u$, and $v$ be two  adjacent vertices in $S$. 
 Then with respect to $S, u, v$, 
 we can partition the vertices in $V(G)\setminus \{u,v\}$ into eight sets as given below, 
 and shown in Figure~\ref{disect}.
\begin{multicols}{2}
\begin{enumerate}[label=(\roman*)]
	\item $N_{S}(uv)=S\cap N(u) \cap N(v)$
	\item $N_{S}{(\bar{u}\bar{v})} = S\cap \overline {N[u]} \cap \overline{N[v]}$ 
	\item $N_S{({u}\bar{v})}=S\cap (N(u)\setminus N[v])$ 
	\item $N_S(\bar{u}{v})=S\cap (N(v)\setminus N[u])$
	\item $N_T(uv)=(N(u)\cap N(v))\setminus S$
	\item $N_T(\bar{u}\bar{v})=(\overline{N[u]}\cap\overline{N[v]})\setminus S$
	\item $N_T({u}\bar{v})=(N(u)\setminus N[v])\setminus S$
	\item $N_T(\bar{u}{v})=(N(v)\setminus N[u])\setminus S$
\end{enumerate}
\end{multicols}
  We notice that $S=N_{S}(uv)\cup N_{S}{(\bar{u}\bar{v})}\cup N_S(u\bar{v})\cup N_S(\bar{u}v)\cup \{u,v\}$.

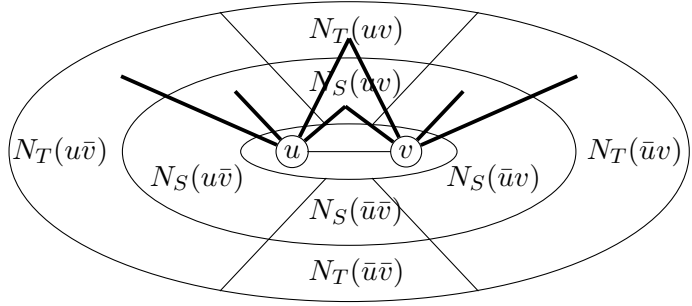
\begin{figure}[ht]
  \centering
    \centering
    \begin{tikzpicture}[myv/.style={ellipse, draw, inner xsep=60pt,inner ysep=25pt}, myv1/.style={ellipse, draw, inner xsep=90pt,inner ysep=40pt},myv2/.style={circle,color=white, draw,inner sep=0.5pt}, myv3/.style={ellipse, draw, inner sep=1.5pt},myv4/.style={circle, draw, inner sep=1.5pt}, myv5/.style={ellipse, draw, inner sep=1.5pt}]

\node[myv2] (a)[label= right:{$N_{S}{(\bar{u}\bar{v})}$}] at (0.3,-0.8) {}; 
\node[myv2] (b)[label= right:{$N_{S}(uv)$}] at (0.3,0.9) {}; 
\node[myv2] (c)[label= below:{$N_S{({u}\bar{v})}$}] at (-1,0) {};
\node[myv2] (d)[label= below:{$N_S(\bar{u}{v})$}] at (2.9,0) {};
\node[myv2] (e)[label= right:{$N_T({u}\bar{v})$}] at (-3.6,0) {};
\node[myv] (f) [label= right:{$N_T(\bar{u}{v})$}] at (1,0) {};

\node[myv2] (h)[label= right:{$N_T(uv)$}] at (0.3,1.6) {};
\node[myv2] (j)[label= right:{$N_T(\bar{u}\bar{v})$}] at (0.3,-1.6) {};

   \node[myv1] (G)[label= right:{}] at (1,0) {};
  \node[myv4] (u) at (0.25,0) {$u$};
  \node[myv4] (v) at (1.75,0) {$v$};
 \node[myv5][fit= (u) (v),  inner xsep=0.25ex, inner ysep=0.25ex] {}; ] {};

 
 
   \draw (0.45,0) -- (1.55,0);
  \draw (-0.7,1.85) -- (0.7,0.35);
  \draw (2.7,1.85) -- (1.3,0.35);
  \draw (0.7,-0.35) -- (-0.7,-1.85);
  \draw (1.3,-0.35) -- (2.7,-1.85);
  \draw (u)[line width=0.5mm] -- (0.95,0.6);
  \draw (u) [line width=0.5mm]-- (-2,1);
  \draw (u) [line width=0.5mm]-- (-0.5,0.8);
  \draw (u) [line width=0.5mm]-- (1,1.5);
  \draw (v)[line width=0.5mm] -- (0.95,0.6);
  \draw (v) [line width=0.5mm]-- (4,1);
  \draw (v) [line width=0.5mm]-- (2.5,0.8);
   \draw (v) [line width=0.5mm]-- (1,1.5);

\end{tikzpicture}
    \caption{Partitioning of vertices of $G$  based on $S$ and two adjacent vertices $u,v \in S$. 
    The bold lines represent the adjacency of  vertices $u$ and $v$~\cite{DBLP:journals/algorithmica/AntonyGPSSS22}.} 
    \label{disect}
\end{figure}

\begin{observation}
\label{ob: not an IS}
Then the following statements are true.
\begin{enumerate}[label=(\roman*)]
    \item $N(u)\setminus N[v]$ induces a $(t-1, t-1)$-split graph 
    with a $(t-1,t-1)$-split partition of ($N_{S}(u\overline{v}), N_{T}(u\overline{v})$).

    \item $N(v)\setminus N[u]$ induces a $(t-1, t-1)$-split graph 
    with a $(t-1,t-1)$-split partition of ($N_{S}(v\overline{u}), N_{T}(v\overline{u})$).

    \item $N_{T}(uv)$ induces an independent set with at most $(t-1)$ vertices.

    \item $N_{S}(\bar{u} \bar{v})$ induces a clique. 
    If $xy$ is an edge of the clique, 
    then $N[x]\cap N[y]$ in $\overline{N[u]} \cap \overline{N[v]}$
    induces a split graph with one split partition being $(N_S(\bar{u}\bar{v}), (N[x]\cap N[y] \cap  \overline{N[u]}\cap \overline{N[v]})\setminus (N_S(\bar{u}\bar{v})))$. 
    
\end{enumerate}
\end{observation}
\begin{proof}
    If $N_S(u\overline{v})$ has a $K_t$, then 
    $v$ along with the vertices of the $K_t$ induce a 
    $K_{1,t}$ in $G\oplus S$. 
    If $N_T(u\overline{v})$ has an independent set of 
    size $t$, then $u$ along with the vertices of the 
    independent set induce a $K_{1,t}$ in $G\oplus S$.
    Therefore, (i) holds true.
    Similarly we can prove the correctness of (ii).
    If there are two adjacent vertices $x$ and $y$ in
    $N_T(uv)$, then $\{x,y,u,v\}$ induces a diamond in
    $G\oplus S$. Therefore, $N_T(uv)$
    is an independent set. If it has at least $t$
    vertices then there is an induced $K_{1,t}$
    formed by those vertices and $u$ in $G\oplus S$.
    Therefore, (iii) holds true.
    If there are two nonadjacent vertices $x$ and $y$
    in $N_S(\bar{u}\bar{v})$,
    then there is a diamond induced by $\{x,y,u,v\}$
    in $G\oplus S$. Therefore, $N_S(\bar{u}\bar{v})$ is a clique. 
    Assume that $x,y\in N_S(\bar{u}\bar{v})$. 
    If $x$ and $y$ have two adjacent common neighbors $x'$ and $y'$ in $N_T(\bar{u}\bar{v})$, then 
    $\{x,y,x',y'\}$ induces a diamond in $G\oplus S$.
    Therefore, $N[x]\cap N[y]\cap \overline{N[u]}\cap \overline{N[v]}$ is a split graph with one split
    partition being $(N_S(\bar{u}\bar{v}), (N[x]\cap N[y] \cap  \overline{N[u]}\cap \overline{N[v]})\setminus (N_S(\bar{u}\bar{v})))$.
    \end{proof}

\begin{lemma}
\label{lem:diamondk1t}
$G$ is a yes-instance of \SCG\ if and only if 
the algorithm returns \YES.
\end{lemma}
\begin{proof}
    Since the algorithm returns \YES\ only when a solution is found, the backward direction of the statement is true. For the forward direction,
    let $G$ be a yes-instance. Assume that 
    there exists a solution $S$ which is an independent set. Further, assume that $G$ has an induced diamond. Then by Lemma~\ref{lem:is-diamond},
    $S$ is the set of all degree-2 vertices of the induced diamonds in $G$. Then Step 1 returns \YES.
    Assume that $G$ is diamond-free. Then by Lemma~\ref{lem:is-k1t}, $S\subseteq I$, where 
    $I$ is the set of isolated vertices in the graph
    induced by the neighbors of $r$, for a center $r$ of an induced 
    $K_{1,t}$ in $G$. Further $|S|\geq |I|-t+2$.
    Then Step 2 returns \YES. 
    Let $S$ be a solution which is not an independent set. Let $uv$ be an edge in the graph induced by $S$.
    The algorithm will discover $uv$ in one iteration of Step 3. 
    By Observation~\ref{ob: not an IS}, we know that 
    the graph induced by
    $N(u)\setminus N[v]$ is a $(t-1,t-1)$-split graph
    with a $(t-1,t-1)$-split partition $(N_S(u\overline{v}), N_T(u\overline{v}))$.
    Similarly, the graph induced by $N(v)\setminus N[v]$
    is a $(t-1,t-1)$-split graph with a 
    $(t-1,t-1)$-split partition $(N_S(\overline{u}v), N_T(\overline{u}v))$. Further, $N_T(uv)$ is an
    independent set of size at most $t-1$. Therefore, in one iteration of Step 3.5, we obtain $S_1=N_S(u\overline{v}), S_2=N_S(\overline{u}v)$, and $S_3=N_S(uv)$. If $N_S(\bar{u}\bar{v})$ is empty,
    then Step 3.5(a) returns \YES. If $N_S(\bar{u}\bar{v})$ is a singleton set,
    then Step 3.5(b) returns \YES. Assume that
    $|N_S(\bar{u}\bar{v})|\geq 2$.
    By Observation~\ref{ob: not an IS}, $N_S(\bar{u}\bar{v})$ is a clique and for every edge $xy$ in it, the common neighborhood of $x$ and $y$ in $\overline{N[u]}\cap \overline{N[v]}$ is a split graph with a partition being $N_S(\bar{u}\bar{v})$ and the rest.
    The algorithm will discover such an edge $xy$
    in one of the iterations of Step 3.5(c) and $N_S(\bar{u}\bar{v})$ will be discovered as $S_4$. Then \YES\ is returned at Step 3.5(c).
\end{proof}

By Proposition~\ref{pro:split-part}, $(t-1,t-1)$-split graphs can be recognized in polynomial-time and all $(t-1,t-1)$-split partitions of a $(t-1,t-1)$-split graph can be found in polynomial-time. Therefore,
each step in the algorithm runs in polynomial-time.
Then we obtain Theorem~\ref{thm:k1tdiamond} from Lemma~\ref{lem:diamondk1t}.
\begin{theorem}
\label{thm:k1tdiamond}
Let $\mathcal{G}$ be the class of $\{K_{1,t},diamond\}$-free graphs for any constant $t\geq 3$. Then \SCG\ can be solved in polynomial-time.
\end{theorem}

It remains open whether the problem is polynomial-time solvable when $\mathcal{G}$ is $H$-free for an $H\in \{K_{1,3},K_{1,4},\text{diamond}\}$.

\bibliographystyle{unsrt}

\bibliography{main}

\end{document}